\documentclass[fleqn]{article}

\usepackage{amsfonts}
\usepackage{amsmath}
\usepackage[amsmath,thmmarks]{ntheorem}
\usepackage[english]{babel}

\usepackage{todonotes}

\newtheorem{theorem}{Theorem}

{\theorembodyfont{\upshape}
	\theoremsymbol{\ensuremath{\Diamond}}
	\newtheorem{definition}[theorem]{Definition}
	
	\theoremsymbol{}
	\newtheorem{example}[theorem]{Example}
	\theoremstyle{nonumberplain}
	\theoremsymbol{}
	\theoremseparator{.}
	\newtheorem{remark}{Remark}
}
\theoremheaderfont{\sc}\theorembodyfont{\upshape}
\theoremstyle{nonumberplain}
\theoremseparator{.}
\theoremsymbol{\rule{1ex}{1ex}}

\usepackage{paralist}
\usepackage{xcolor}
\usepackage{xspace}
\usepackage{tikz}
\usetikzlibrary{quantikz2}
\usepackage{xspace}
\usepackage{url}
\usepackage[utf8]{inputenc}

\newcommand{\isdef}{\stackrel{\mbox{\tiny def}}{=}}

\newcommand{\set}[1]{\left\{#1\right\}}
\newcommand{\tuple}[1]{\langle #1 \rangle}
\newcommand{\card}[1]{\left\vert#1\right\vert}
\newcommand{\norm}[1]{\left\lVert#1\right\rVert}
\newcommand{\setof}[2]{\left\{#1\,\middle|\:#2\right\}}

\newcommand{\dN}{\mathbb{N}}

\newcommand{\dR}{\mathbb{R}}
\newcommand{\dB}{\mathbb{B}}

\newcommand{\dH}{\mathcal{H}}
\newcommand{\dC}{\mathbb{C}}
\newcommand{\dE}{\mathbb{E}}

\newcommand{\calA}{\mathcal{A}}
\newcommand{\calM}{\mathcal{M}}
\newcommand{\calN}{\mathcal{N}}

\newcommand{\interv}[2]{[#1,#2]}

\newcommand{\inner}[2]{\langle #1|#2\rangle}
\newcommand{\outerp}[2]{\ket{#1}\!\bra{#2}}
\newcommand{\conj}[1]{\overline{#1}}
\newcommand{\adj}[1]{#1^\dagger}
\newcommand{\vzero}{\mathbf{0}}
\newcommand{\idty}{\mathbf{I}}
\newcommand{\trc}[1]{\mathrm{Tr}\left(#1\right)}
\newcommand{\cmat}[2]{\dC^{#1\times #2}}

\newcommand{\lmspace}{\vspace{0.5em}}
\newcommand{\aindent}{\quad}
\newcommand{\sindent}{}

\newcommand{\isatext}[1]{\textcolor{blue}{{#1}}}
\newcommand{\keyw}[1]{\textcolor{MyGreen}{\textbf{#1}}}
\newcommand{\isacmd}[1]{\textcolor{MyPurple}{\texttt{#1}}}
\newcommand{\isalem}[1]{\textcolor{MyPurple}{\textsc{#1}}}
\newcommand{\lhv}{\isacmd{lhv}}
\newcommand{\msr}{\isacmd{measure}}
\newcommand{\pbspace}{\isacmd{prob-space}}
\newcommand{\bmeas}{\isacmd{borel-measurable}}
\newcommand{\posrv}{\isacmd{pos-rv}}
\newcommand{\psrv}{\isacmd{prv-sum}}
\newcommand{\fcmat}{\isacmd{fixed-carrier-mat}}
\newcommand{\dimr}{\isacmd{dimR}}
\newcommand{\dimc}{\isacmd{dimC}}
\newcommand{\fcmats}{\isacmd{fc-mats}}
\newcommand{\summat}{\isacmd{sum-mat}}
\newcommand{\sumwith}{\isacmd{sum-with}}
\newcommand{\isaset}{\isacmd{set}}
\newcommand{\cpxsqmat}{\isacmd{cpx-sq-mat}}
\newcommand{\carrier}{\isacmd{carrier-mat}}
\newcommand{\gmat}{\isacmd{mat}}
\newcommand{\measurable}{\isacmd{measurable}}
\newcommand{\sepdensity}{\isacmd{separable-density}}
\newcommand{\prjct}[1]{\Pi^{#1}}
\newcommand{\ltwo}{\mathcal{L}_2}
\newcommand{\nrm}[1]{\|#1\|_2}
\newcommand{\cmod}[1]{\card{#1}}
\newcommand{\commut}[2]{[#1,#2]}
\newcommand{\cpxmat}{\isacmd{complex }\gmat}
\newcommand{\spct}[1]{\mathrm{spct}(#1)}
\newcommand{\projector}{\isacmd{projector}}
\newcommand{\dimrow}{\isacmd{dim-row}}
\newcommand{\hermitian}{\isacmd{hermitian}}
\newcommand{\unitary}{\isacmd{unitary}}
\newcommand{\AEv}{\textsc{AE}}
\newcommand{\probspace}{\isacmd{prob-space}}
\newcommand{\finitemeasure}{\isacmd{finite-measure}}
\newcommand{\ieth}{\mathrm{th}}
\newcommand{\transp}{\mathrm{T}}
\newcommand{\calL}{\mathcal{L}}
\newcommand{\opnrm}[1]{\|#1\|_{\mathrm{op}}}
\newcommand{\singval}[1]{\sigma^*(#1)}

\newenvironment{defisa}{\[\begingroup\color{blue}\begin{array}{lcl}}{\end{array}\endgroup\]}
\newenvironment{lemisa}{\[\begingroup\color{blue}\begin{array}{l}}{\end{array}\endgroup\]}
\definecolor{MyPurple}{RGB}{7,17,107}
\definecolor{MyGreen}{RGB}{7,107,10}

\begin{document}
	\title{A formalization of the CHSH inequality and Tsirelson's upper-bound in Isabelle/HOL}
	
	\author{Mnacho Echenim and Mehdi Mhalla\\ {Universit\'e Grenoble Alpes, Grenoble INP\footnote{Institute of Engineering Univ. Grenoble Alpes}}\\ {CNRS, LIG, F-38000 Grenoble, France}
	}
	
	\date{June 2023}
	

	\maketitle
	\begin{abstract}
		We present a formalization of several fundamental notions and results from Quantum Information theory, including density matrices and projective measurements, along with the proof that the local hidden-variable hypothesis advocated by Einstein to model quantum mechanics cannot hold. The proof of the latter result is based on the so-called CHSH inequality, and it is the violation of this inequality that was experimentally evidenced by Aspect who earned the Nobel Prize in 2022 for his work. We prove various results related to the violation of the CHSH inequality, such as Tsirelson's bound which permits to obtain the maximum violation of this inequality in a quantum setting.
	\end{abstract}




\section{Introduction}

%
%

The main goal of a proof assistant is to allow its user to produce a certified mathematical proof, i.e., a proof that is irrefutable. The reason it can be necessary to certify a proof is that almost all proofs found in articles or textbooks do not spell out every reasoning step -- this would make the entire proof unreadable --, and there can be errors in these gaps that go undetected for a long time, especially in complicated proofs. This phenomenon has been particularly manifest in Quantum Information theory, a domain in which erroneous proofs have had significant impacts:
\begin{itemize}
	\item A major result in computability theory was published in \cite{ji2021mip}, which states that (classical) recursively enumerable languages  are exactly the languages that can be decided by a classical polynomial-time verifier interacting with several quantum provers sharing an entangled state.The original proof relied on a result from \cite{vidick2016three} that turned out to be wrong. The computability proof was fixed by getting rid of the dependency on the result from \cite{vidick2016three}, and it is currently still unknown whether the latter is true or not.
	\item It was recently discovered \cite{berta2022gap} that there is an error in the proof of a result called the \emph{generalized quantum Stein's lemma} \cite{brandao2010generalization} which 
	makes it unclear whether several other results related to the reversibility of quantum entanglement and general quantum resources that are based on this lemma (such as, e.g., \cite{brandao2010reversible, brandao2008entanglement, brandao2015reversible} that together have over 700 citations) are correct or not.
\end{itemize}
The case can thus be made that being able to certify results in quantum computation and information is an important task, probably more so because of the many counter-intuitive results that have been derived in the field.

In this paper we present the formalization in Isabelle/HOL of several results related to the \emph{CHSH inequality} \cite{CHSH}. This is one of the Bell inequalities, which are related to two of the postulates of quantum mechanics: 
\begin{inparaenum}[i)]
	\item the measurement postulate which describes the probabilistic outcomes of measuring a quantum system and the way this system changes due to measurements; and
	\item the composite state postulate, which entails the existence of \emph{entangled} particles, i.e., distinct particles that remain correlated regardless of the distance between them.
\end{inparaenum}
We formalize the CHSH inequality and the proof of its violation in a quantum setting along with three related results:
\begin{itemize}
	\item When the quantum system that is measured is not entangled, the CHSH inequality cannot be violated.
	\item When the measuring devises are not appropriately chosen (when the related observables commute), the CHSH inequality cannot be violated.
	\item The maximum violation of the inequality in a quantum setting is $2\cdot\sqrt{2}$, a result known as Tsirelson's upper-bound \cite{Tsirelson}.
	\item Tsirelson's upper-bound is tight and can be reached by an appropriate selection of quantum states and measurement devices.
\end{itemize}

\subsubsection*{On the EPR paradox and Bell inequalities}
The fact that a physical system can be in a superposition of states and that, instead of revealing a pre-existing value, a measurement ``brings the outcome into being'' (\cite{Mermin93}) was the cause of many controversies between the pioneers of quantum mechanics. Famously, Einstein did not believe in the intrinsically statistical nature of quantum mechanics. According to him, quantum mechanics was an incomplete theory, and the postulates on probabilistic measure outcomes actually reflected statistical outcomes of a deterministic underlying theory (see \cite{Dalton20,scarani2019bell} for detailed considerations on these controversies). The EPR paradox \cite{EPR} was designed to evidence the incompleteness of quantum mechanics. 
It involves two distinct particles that are entangled -- a physical phenomenon that intuitively connects both particles -- and which are sent in opposite directions. If one of the particles is measured, then the outcome of the measurement of the other particle will be known with certainty. 
This  phenomenon is known as \emph{nonlocality}, and it may give the impression  that information traveled from the first to the second particle instantaneously, which would 
contradict the theory of relativity\footnote{Since then, it has been proven that this phenomenon is in no contradiction with the theory of relativity and does not imply faster-than-light communication.}. Einstein called this phenomenon ``spooky action at a distance''. A suggested solution to this phenomenon is that the measurement outcomes are actually properties that existed \emph{before} the measurement was performed, and that deterministic underlying theories for quantum mechanics should thus be developed.
Efforts to develop such theories are called \emph{hidden-variable programs}. The theories that also take into account the fact that information cannot travel instantaneously, thus also requiring that distant events are independent, are called \emph{local hidden-variable theories}. 

The fact that there can be no local hidden-variable underlying theory for quantum mechanics was proved by Bell \cite{Bell64} who derived inequalities (the \emph{Bell inequalities}), that hold in a probabilistic setting, and showed that they are violated by measurements in quantum mechanics. Since his seminal work, other Bell inequalities that hold in a probabilistic setting but are violated in the quantum setting have been derived, among which the CHSH inequality, named after Clauser, Horne, Shimony and Holt \cite{CHSH}. This is the inequality that was experimentally violated by Aspect \cite{aspect}, who, along with Clauser and Zeilinger, was awarded the Nobel Prize in Physics for his his work on entangled particles. 

\subsubsection*{Related work}
There are several recent or ongoing lines of research on the use of formal tools for the analysis of quantum algorithms and protocols, such as the formal verification framework for quantum programs \textsc{Qbricks} \cite{CharetonBBPV21}, an extension of attack trees with probabilities \cite{Kammuller19}, or the development of dedicated quantum Hoare logics \cite{Unruh_2019,liu19}. Approaches that are closer to ours involve the formalization of quantum notions and algorithms in proof assistants including Coq \cite{Boender_2015,Coq_Quantum} and Isabelle \cite{isa-dirac}. The formalization work described in \cite{isa-dirac} provides an example of the importance of certifying results in Quantum Information. As noted in their paper, they uncovered an error in a highly cited article that had been published in a high-level physics journal as they attempted to formalize the corresponding result.
In this paper we extend the effort started in \cite{isa-dirac} in two ways. First, our formalization is based on so-called \emph{density operators} rather than pure quantum states representing states in a Hilbert space of an arbitrary dimension. Density operators are a more convenient way of representing quantum systems that are in a mixed state, i.e., in one of several pure quantum states with associated probabilities that sum to 1. They also permit to obtain more general and natural statements on quantum mechanics. Second, although notions related to measurements are formalized in \cite{isa-dirac}, these are specific and in particular, only involve measurements in the standard basis. We present the full formalization of projective measurements and observables in this paper. To the best of our knowledge, no such formalization is available in any proof assistant. The main formalization of notions around the CHSH inequality is available in the Archive of Formal Proofs \cite{TsirelsonAFP}, it relies on our formalization of projective measurements available in \cite{ProjectiveAFP}. 

\subsubsection*{Organization of the paper}
The paper is organized as follows. Section \ref{sec:prelim} contains an overview of Isabelle/HOL, along with some notations and results about measure theory that are available in the distribution of Isabelle and are used for the formalization of the local hidden-variable hypothesis. The section also introduces concepts from linear algebra along with their notations in a quantum setting. Section \ref{sec:postulates} is devoted to the presentation of the postulates of quantum mechanics, and we present the CHSH inequality and the formalization of the related results in Section \ref{sec:chsh}. We mention future research directions in Section \ref{sec:conclusion}.
\section{Preliminaries}\label{sec:prelim}

We review the formalization of probability theory in Isabelle, as well as  notions from linear algebra that will be used in the formalization of quantum mechanics. Our formalization of this latter topic is mainly based on \cite{mathQuantum, nielsen-book}. As we will only formalize quantum notions in finite dimension, we present the standard general definitions and illustrate them in the finite dimensional case. For the sake of readability, we chose to omit trivial hypotheses in the statements of the lemmas given below. For example, we did not add in statements of lemmas involving matrices that the formalized results hold for matrices with a nonzero number of lines and columns.

\subsection{Isabelle/HOL}
Our formalization was carried out in the interactive theorem prover Isabelle/HOL. This tool can be downloaded at \url{https://isabelle.in.tum.de/}, along with tutorials and documentations; additional material on Isabelle can be found in \cite{ConcreteSemantics}. This prover is based on the typed $\lambda$-calculus. Terms are built using types that can be: 
\begin{itemize}
	\item simple types, denoted with the Greek letters $\alpha, \beta,\ldots$
	\item types obtained from type constructors, represented in postfix notation (e.g. the type \isatext{$\alpha$ \isacmd{set}} which denotes the type of sets containing elements of type $\alpha$), or in infix notation (e.g., the type \isatext{$\alpha\rightarrow \beta$} which denotes the type of total functions from $\alpha$ to $\beta$).
\end{itemize}
Functions are curried, and function application is written without parentheses. Anonymous functions are represented with the lambda notation: the function $x\mapsto t$ is denoted by \isatext{$\lambda x.\,t$}. We will use  mathematical notations for standard terms; for example, the set of reals will be denoted by $\dR$ and the set of booleans by $\dB$. The application of function $f$ to argument $x$ may be written \isatext{$f\ x$}, \isatext{$f(x)$} or \isatext{$f_x$} for readability.

This work relies on Isabelle's mechanism of \emph{locales} \cite{Ballarin14}. Intuitively, a locale represents a proof context consisting of parameters and assumptions. Locales are simple to define and combine into hierarchies. In our opinion, using locales permits to formalize notions in a  natural way thanks to their features, and there is a lot of evidence that their usage permits to formalize complex mathematics in Isabelle/HOL's simple type theory, see, e.g., \cite{Bordg22}.


\subsection{Probability theory in Isabelle/HOL}

A large part of the formalization of measure and probability theory in Isabelle was carried out in \cite{hoelzl2012thesis} and is  included in Isabelle's distribution.  We briefly recap some of the notions that will be used throughout the paper and the way they are formalized in Isabelle. We assume the reader has knowledge of fundamental concepts of measure and probability theory; any missing notions can be found in 
\cite{Durrett} for example. 
Probability spaces are particular \emph{measure spaces}.  A measure space over a set $\Omega$ consists of a function $\mu$ that associates a nonnegative number {or $+\infty$} to  some subsets of $\Omega$. {The subsets of $\Omega$ that can be measured are closed under  complement and countable unions and make up a \emph{$\sigma$-algebra}.} In Isabelle the measure type with elements of type $\alpha$ is denoted by $\alpha\ \msr$.
A function between two measurable spaces is \emph{measurable} if the preimage of every measurable set is measurable. 
In Isabelle, sets of measurable functions are defined as follow:
\begin{defisa}
	\measurable & :: & \alpha\,\msr \rightarrow \beta\,\msr\rightarrow \left(\alpha \rightarrow \beta\right) \isaset\\
	\measurable\ \calM\ \calN\ \mu & =&
	\setof{f: \Omega_\calM \rightarrow \Omega_\calN}{\forall A\in \calA_\calN.\, f^{-1}(A) \cap \Omega_\calM \in \calA_M}
\end{defisa}
Measurable functions that map the elements of a measurable space into real numbers, such as  random variables which are defined below, are measurable on Borel sets:
\begin{lemisa}
	\keyw{abbreviation}\ \bmeas\ \calM\ \equiv\  \measurable\ \calM\ \isacmd{borel}  
\end{lemisa}	
Probability measures are measure spaces on which the measure of $\Omega$ is finite and equal to 
$1$. In Isabelle, they are defined in a {locale}:
\begin{lemisa}
	\keyw{locale}\ \probspace =  \finitemeasure\ +\keyw{ assumes}\ \mu_\calM(\Omega_\calM)\ =\ 1 
\end{lemisa}
A \emph{random variable} on a probability space $\calM$ is a measurable function with domain $\Omega_\calM$. The average value of a random variable $f$ is called its \emph{expectation}, it is denoted by\footnote{The superscript is omitted when there is no confusion.} $\dE^\calM[f]$, and defined by $\dE^\calM[f] \isdef \int_{\Omega_\calM}f\mathrm{d}\mu_\calM$.

In what follows, we will consider properties that hold \emph{almost surely} (or \emph{almost everywhere}), {i.e.,} are such that the elements for which they do not hold reside within a set of measure $0$:
\begin{lemisa}
	\keyw{lemma}\ \isalem{AE-iff}:\\
	\sindent \keyw{shows}\ (\AEv_\calM\,x.\ P\ x) \Leftrightarrow (\exists N\in \calA_\calM.\, \mu_\calM(N) = 0 \wedge \setof{x}{\neg P\ x}\subseteq N)
\end{lemisa}
We will formalize results involving local hidden-variable hypotheses under the more general assumption that properties hold almost everywhere, rather than on the entire probability space under consideration.

%
%
\subsection{Linear algebra formalizations}

We recap the core notions from linear algebra that will be used in this work. Some of these notions had already been formalized in \cite{QHLProver,afp-isa-dirac}, building over the work in \cite{jordan_afp}. A  detailed treatment on linear algebra can be found, e.g., in \cite{linalg}.

\paragraph{On Hilbert spaces.}
A \emph{Hilbert space} $\dH$ is a complete vector space over the field of complex numbers, equipped with an \emph{inner product} $\inner{\cdot}{\cdot}: \dH \times \dH \rightarrow \dC$, i.e., a function such that for all $\varphi, \psi\in \dH$, $\inner{\varphi}{\psi} = \conj{\inner{\psi}{\varphi}}$, $\inner{\varphi}{\varphi} \geq 0$ and $\inner{\varphi}{\varphi} = 0$ if and only if $\varphi = \vzero$. The \emph{norm} induced by the inner product is defined by $\norm{\varphi} \isdef \sqrt{\inner{\varphi}{\varphi}}$, and $\varphi$ is \emph{normalized} if $\norm{\varphi} = 1$. We assume all Hilbert spaces under consideration are finite dimensional. The elements of a Hilbert space of dimension $n$ are represented as column vectors, and the inner product of $\varphi$ and $\psi$ is $\inner{\varphi}{\psi} \isdef \sum_{i = 1}^n \conj{\varphi_i}\psi_i$. The definitions and core properties related to inner products and induced norms are formalized in Isabelle in \cite{QHLProver}. 	

It is standard in quantum mechanics to represent the elements of $\dH$ using the Dirac notation $\ket{\cdot}$, so that vector $u$ is denoted by $\ket{u}$. The elements in $\dH$ are sometimes called \emph{ket-vectors}. In what follows, we will use the standard mathematical terminology for vectors to define the preliminary notions, and we will use the ket notation in the quantum setting.
\begin{example}\label{ex:ket}
	Given a two-dimensional Hilbert space $\dH$, of which the vectors $\begin{pmatrix} 1\\ 0 \end{pmatrix}$ and $\begin{pmatrix} 0\\ 1 \end{pmatrix}$ form a basis, it is standard to denote these vectors as follows:
	\[\ket{0}\ \isdef\ \begin{pmatrix} 1\\ 0 \end{pmatrix} \quad \text{and}\quad \ket{1}\ \isdef\ \begin{pmatrix} 0\\ 1 \end{pmatrix}.\]
	The set $\set{\ket{0}, \ket{1}}$ is called the \emph{standard basis} of $\dH$. 
	
	Two other common ket-vectors are $\ket{+} \isdef \frac{1}{\sqrt{2}}\cdot\left(\ket{0} + \ket{1}\right)$ and $\ket{-}\isdef \frac{1}{\sqrt{2}}\cdot\left(\ket{0} - \ket{1}\right)$. Note that $\ket{0}$, $\ket{1}$, $\ket{+}$ and $\ket{-}$ are all normalized.
\end{example}

An \emph{operator} is a linear map on a Hilbert space, and we will identify operators with their matrix representations. The set of matrices with $n$ rows and $m$ columns is represented in Isabelle by \isatext{$\carrier\ n\ m$}, and the type of complex matrices is represented in Isabelle by \isatext{\cpxmat}. We denote by $\cmat{n}{m}$ the set of complex matrices with $n$ rows and $m$ columns and by $\vzero_{n,m}$ the matrix in $\cmat{n}{m}$ containing only zeroes. When there is no ambiguity, we may write $\vzero$ instead of $\vzero_{n,m}$. We denote by $\idty$ the identity operator; we may also write $\idty_n$ to specify that the considered Hilbert space is of dimension $n$. The \emph{trace} of a square matrix is the sum of its diagonal elements: if $A\in \cmat{n}{n}$ then $\trc{A} = \sum_{i=1}^n A_{i,i}$. 

In the quantum mechanical setting, it will often be necessary to work with general sums of matrices over finite sets, but such an operation cannot be defined on the entire set of matrices, since their dimensions may differ.
In order to still handle such sums and apply the useful lemmas that have been already formalized in a general setting, we used the types-to-sets tool \cite{types-sets} that permits to transform type-based statements which hold on an entire type universe to their set-based counterparts. This tool is particularly useful to apply lemmata in contexts where the assumptions only hold on a strict subset of the considered universe. For example, generalized summation over sets is defined for types that represent abelian semigroups with a neutral element. Such an algebraic structure is straightforward to define on any set of matrices that all have the same  dimensions. In our formalization, we use a locale to only consider the set of matrices in which the number of rows and columns is fixed:
\begin{lemisa}
	\keyw{locale}\ \fcmat\ =\  \keyw{fixes}\ \fcmats\ \dimr\ \dimc\\
	\aindent\keyw{assumes}\ \fcmats\ =\ \carrier\ \dimr\ \dimc 
\end{lemisa}
All matrices in {\fcmats} thus admit $\dimr$ rows and $\dimc$ columns. After proving that $\fcmats$ along with the standard addition on matrices and the matrices with $\dimr$ rows and $\dimc$ columns consisting only of zeroes is an abelian semigroup with a neutral element, we can define a generalized summation of matrices on this locale:
\begin{defisa}
	\summat & :: & (\alpha \rightarrow \beta\ \gmat) \rightarrow \alpha\ \isaset\ \rightarrow \beta\ \gmat\\
	\summat\ \calA\ I &= & \sumwith\ (+)\ \vzero\ \calA\ I
\end{defisa}
The types-to-sets transfer tool then permits with no effort to transfer theorems that hold on abelian semigroups, and especially those involving sums, to this locale.	
As all the matrices we consider in what follows are nontrivial complex square matrices, we extend the previous locale to work in this context:
\begin{lemisa}
	\keyw{locale}\ \cpxsqmat\ =\  \fcmat\ (\fcmats:: \cpxmat\ \isaset)\, +\\
	\aindent\keyw{assumes}\ \dimr = \dimc\ \keyw{and}\ \dimr > 0
\end{lemisa}
We will consider the  notion of \emph{bra-vectors}, denoted by $\bra{u}$. Formally, the bra-vector $\bra{u}$ is the linear map that maps the vector $\ket{v}$ to the complex number $\inner{u}{v}$. In the finite-dimensional setting, we have $\bra{u} = (\conj{u_1},  \ldots, \conj{u_n})$ when \[\ket{u} = \begin{pmatrix}
	u_{1} \\
	\vdots \\
	u_{n}
\end{pmatrix}.\]
By a slight abuse of notation, we will identify the application of $\bra{u}$ to $\ket{v}$ with the inner product $\inner{u}{v}$.
Given two vectors $\ket{u}$ and $\ket{v}$, we define their \emph{outer product}, denoted by $\outerp{u}{v}$, as the linear map such that, for all $\ket{v'} \in \dH$, $(\outerp{u}{v})\ket{v'} = \ket{u}\cdot\inner{v}{v'} = \inner{v}{v'}\cdot \ket{u}$. In the finite-dimensional setting, this outer product is represented by the matrix $M$, where $M_{i,j} = u_i\cdot \conj{v_j}$.	The outer product $\outerp{v}{v}$ is called the \emph{rank one projector on $v$}. 
\begin{example}
	Following Example \ref{ex:ket}, we have the following expressions for the outer products $\outerp{-}{1}$  and $\outerp{-}{-}$:
	\[\outerp{-}{1}\ =\ \frac{1}{\sqrt{2}}\cdot \begin{pmatrix}
		0 & 1\\
		0 & -1
	\end{pmatrix} \quad \text{and} \quad \outerp{-}{-}\ =\ \frac{1}{2}\cdot \begin{pmatrix}
		1 & -1\\
		-1 & 1	
	\end{pmatrix}.\]
	Outer products and inner products are related by the equation $(\outerp{u}{v}) \cdot (\outerp{w}{x}) = \inner{v}{w}\cdot(\outerp{u}{x})$ (Lemma \isalem{outer-prod-mult-outer-prod} in \cite{QHLProver}). Using this equation, we  have 
	\[(\outerp{-}{1})\cdot \left(\outerp{-}{-}\right) = \inner{1}{-}\cdot( \outerp{-}{-}) = -\frac{1}{\sqrt{2}}\cdot (\outerp{-}{-}) = -\frac{1}{2\sqrt{2}}\cdot \begin{pmatrix}
		1 & -1\\
		-1 & 1	
	\end{pmatrix}.\]
\end{example}

\paragraph{On spectral decomposition.}
We introduce the notions of projectors, and Hermitian and unitary operators. These are essential in the definition of quantum mechanics postulates. If $A$ is an operator on $\dH$, then the operator $B$ such that for all $\varphi, \psi \in \dH$ $\inner{\varphi}{A\psi} = \inner{B\varphi}{\psi}$ is called the \emph{adjoint of $A$}, and denoted by $\adj{A}$. For all matrices $A$ and $B$, we have $\adj{(A\cdot B)} = \adj{B}\cdot \adj{A}$. Note that we also have $\adj{(\ket{u})} = \bra{u}$. If $A = \adj{A}$ then we say that $A$ is \emph{self-adjoint} or \emph{Hermitian},  and if $A\cdot\adj{A} = \adj{A}\cdot A = \idty$, then we say that $A$ is \emph{unitary}.   
An operator $A$ is a \emph{projector} if $A^2 = A$, and $A$ is an \emph{orthogonal projection} if $\adj{A} = A$. We say that $A$ is \emph{positive} if, for all $\varphi \in \dH$, we have $\inner{\varphi}{A\varphi} \geq 0$. 	
We say that $\varphi$ is an \emph{eigenvector} of operator $A$ is there exists $a\in \dC$ such that $A\cdot\varphi = a\cdot \varphi$. In this case, we say that $a$ is an \emph{eigenvalue} of $A$. The set of eigenvalues of $A$ is called the \emph{spectrum} of $A$ and denoted by $\spct{A}$. 
When $A$ is Hermitian, all its eigenvalues are necessarily real and it is possible to associate every eigenvalue $a \in \spct{A}$ with a projector $\prjct{a}$ such that
\[\prjct{a}\cdot \prjct{a'} = \vzero\ \text{if}\ a \neq a',\ \sum_{a \in \spct{A}}\prjct{a} = \idty\ \text{and}\ A\ =\ \sum_{a \in \spct{A}}a\cdot \prjct{a}.\]
These properties are at the core of the quantum measurement postulate introduced below. They are obtained in our formalization by introducing a predicate stating that a matrix can be decomposed as a product involving a unitary matrix and   a diagonal one containing only real elements:
\newcommand{\rdiagdec}{\isacmd{real-diag-decomp}}
\newcommand{\diagmat}{\isacmd{diagonal-mat}}
\begin{defisa}
	\rdiagdec & :: & \cpxmat \rightarrow \cpxmat \rightarrow \cpxmat \rightarrow \dB\\
	\rdiagdec\ A\ B\ U &\equiv & A = U\cdot B\cdot U^\dagger \wedge \unitary\ U \wedge \diagmat\ B \wedge\\
	& & \quad \forall i < \dimrow\ B.\ B_{i,i} \in \dR
\end{defisa}
The following lemma states that hermitian matrices always admit such a decomposition:
\begin{lemisa}
	\keyw{lemma}\ \isalem{hermitian-real-diag-decomp}:\\
	\aindent \keyw{assumes}\ \hermitian\ A\\
	\sindent \keyw{obtains}\ B\ U\ \keyw{where}\ \rdiagdec\ A\ B\ U
\end{lemisa}
The proof of this lemma is based on the Schur decomposition, which was formalized in \cite{jordan} and refined in \cite{QHLProver}. The decomposition above admits the following key properties:
\begin{itemize}
	\item The spectrum of $A$ consists of the set of elements on the diagonal of $B$ (Lemma \isalem{unitary-diag-spectrum-eq}).
	\item The columns of $U$ are pairwise orthogonal (Lemma \isalem{unitary-col-inner-prod}).
	\item If $U_i$ denotes the $i^{\ieth}$ column of $U$ then we have
	\[A = \sum_{i=1}^n B_{i,i}\cdot\outerp{U_i}{U_i}\quad \text{(Lemma \isalem{sum-decomp-cols})}.\]
\end{itemize}

\paragraph{Tensor products.}
We present the formalization of \emph{tensor products} that will be used in this paper. From an abstract point of view, a tensor product is a bilinear map, but the formalization we consider is based on the Kronecker product and was developed independently in \cite{QHLProver} and \cite{afp-isa-dirac}. It is standard to use the symbol $\otimes$ to denote tensor products. We stick to this convention: the tensor product of	
two vector spaces $U$ and $V$ is denoted by $U\otimes V$; the tensor product of vectors $u\in U$ and $v\in V$ is denoted by $u\otimes v$ and the tensor product of matrices $A$ and $B$ is denoted by $A\otimes B$. In the quantum setting, when convenient, we may write $\ket{uv}$ instead of $\ket{u}\otimes \ket{v}$ and similarly, we may write $\bra{uv}$ instead of $\bra{u}\otimes \bra{v}$.

\begin{example}
	We have the following:
	\[\begin{array}{rclcl}
		\ket{0}\otimes \ket{0} & = & \ket{00} & = & (1\ 0\ 0\ 0)^\transp,\\
		\ket{0}\otimes \ket{1} & = & \ket{01} & = & (0\ 1\ 0\ 0)^\transp,\\
		\ket{1}\otimes \ket{0} & = & \ket{10} & = & (0\ 0\ 1\ 0)^\transp,\\
		\ket{1}\otimes \ket{1} & = & \ket{11} & = & (0\ 0\ 0\ 1)^\transp.
	\end{array}\]
	In particular if $\set{\ket{0}, \ket{1}}$ is a basis for  $\dH$, then $\set{\ket{00},\ket{01},\ket{10},\ket{11}}$ is a basis for $\dH\otimes \dH$.
\end{example}

\paragraph{The $\calL_2$ operator norm.}
There are several equivalent definitions of the $\calL_2$ operator norm of a matrix $A$, our formalization is based on the following definition:
$\opnrm{A}\ \isdef\ \sup \setof{\nrm{A\cdot v}}{\nrm{v} = 1}$. When this notion was formalized, we ran into an issue because when $v$ is a complex vector, $\nrm{v}$ is a complex number in Isabelle. This makes definitions and proofs difficult to work with, for example, the supremum of a set of complex numbers is not defined. We found that the simplest way to overcome this issue was to slightly modify the definition by replacing $\nrm{v}$ by its real part $\mathrm{Re}\ \nrm{v}$, in order to constrain the type of the considered set to be real. 

In the finite dimensional case, the $\calL_2$ operator norm of a matrix $A$ can be determined by  the \emph{singular values} of $A$, i.e., the square roots of the eigenvalues of $\adj{A}\cdot A$. If $\singval{A}$ denotes the maximum singular value of $A$, then $\opnrm{A} = \singval{A}$ (Lemma \isalem{L2-op-nrm-max-sgval-eq}). 

\newcommand{\rankproj}{\isacmd{rank-1-proj}}
\newcommand{\adjoint}{\isacmd{adjoint}}

\section{The postulates of quantum mechanics}\label{sec:postulates}
We introduce the postulates of quantum mechanics following \cite{nielsen-book} and detail the way the related notions are formalized in Isabelle, as well as some of the choices in the formalization process.
\subsection{Single and composite states}
\paragraph{The state postulate.} The state postulate is the following: 
\begin{quote}
	\emph{Associated with an isolated physical system is a Hilbert space, which is referred to as the \emph{state space}. The state of the system itself is completely described by its density operator, i.e., a positive, semi-definite Hermitian operator with trace 1 that acts on the underlying Hilbert space of the physical system.}
\end{quote}
In the finite-dimensional setting, such an operator is represented by a \emph{density matrix}, and it is not uncommon to use both terms interchangeably. This notion has already been formalized in \cite{liu19}: 
\newcommand{\densop}{\isacmd{density-operator}}
\begin{defisa}
	\densop & :: & \cpxmat \rightarrow \dB\\
	\densop\ \rho &\equiv & \isacmd{positive}\ \rho\ \wedge\ \trc{\rho} = 1
\end{defisa}
It is based on the fact that every positive matrix is necessarily Hermitian. When the Hilbert space is of dimension two, the physical system is called a \emph{qubit}.
\begin{example}\label{ex:basic:density}
	The rank one projectors on the first and second basis state of a Hilbert space with dimension two are respectively
	\[\rho_0 \isdef \outerp{0}{0} = \begin{pmatrix}
		1 & 0\\
		0 & 0
	\end{pmatrix}\quad \text{ and } \quad \rho_1 \isdef \outerp{1}{1} = \begin{pmatrix}
		0 & 0\\
		0 & 1
	\end{pmatrix};\] 
	these are both density matrices. 
	Given  $\alpha,\beta \in \dR$ such that $0 \leq \alpha,\beta$ and $\alpha + \beta = 1$,  $\alpha\cdot \rho_0 + \beta\cdot \rho_1$ is  a also density matrix. For example, $\frac{1}{2}\cdot\rho_0+\frac{1}{2}\cdot \rho_1=\frac{1}{2}\cdot\idty_2$ is a density matrix. Such density matrices represent states that are built by mixing different states with  coefficients that can be viewed as  probabilities, they are called \emph{mixed states}.
\end{example}

\begin{example}\label{ex:superposition}
	Given a Hilbert space $\dH$ of dimension 2 and vector $\ket{\varphi} = \alpha\cdot \ket{0} + \beta\cdot \ket{1} \in \dH$ where $\alpha, \beta\in \dC$ and $\cmod{\alpha}^2 + \cmod{\beta}^2 = 1$,  the projector  $\outerp{\varphi}{\varphi}$ 
	is a density matrix. 
	It is important to note that in general, such density matrices admit off-diagonal elements and are therefore different from the mixtures of basis states defined in Example \ref{ex:basic:density}: when this is the case, the considered density matrix represents a physical system in a \emph{superposition state}. 
\end{example}

In the general case, density matrices are exactly the matrices that are of the form $\sum_j \lambda_j\cdot\outerp{\psi_j}{\psi_j}$, where the $\lambda_j$s are nonnegative numbers that sum to 1 and each $\psi_j$ has norm 1: 

\begin{lemisa}
	\keyw{lemma}\ \isalem{density-operator-iff-mixed-state}:\\
	\aindent \keyw{assumes}\ \rho \in \fcmats\\
	\sindent \keyw{shows}\ (\densop\ \rho) \longleftrightarrow \\
	\sindent\ \ (\exists p\ \psi\ n.\  (\forall i < n.\  0 \leq \lambda_i)\ \wedge\ (\forall i < n. \norm{\psi_i} = 1)\ \wedge\ (\sum_{i<n} \lambda_i = 1)\ \wedge\ \\
	\sindent\quad\ \rho = \sum_{i<n} \lambda_i \cdot \outerp{\psi_i}{\psi_i})
\end{lemisa}

\begin{remark}
	Note that distinct ensembles can be associated with the same density matrix. For example, the ensembles  
	\[\set{\left(\frac{1}{2}, \outerp{0}{0}\right),\ \left(\frac{1}{2}, \outerp{1}{1}\right)} \quad\text{and}\quad \set{\left(\frac{1}{2}, \outerp{+}{+}\right),\ \left(\frac{1}{2}, \outerp{-}{-}\right)}\]
	are both associated with the density matrix $\frac{1}{2}\cdot \idty_2$. As all the quantum operations are defined on density matrices, this entails that these ensembles are indistinguishable.
\end{remark}

Most introductory textbooks on Quantum Information theory provide a version of the state postulate that is based on so-called state vectors $\ket{\psi}$. The latter represent \emph{pure states}, for which the corresponding density matrix is $\outerp{\psi}{\psi}$:
\newcommand{\puredensop}{\isacmd{pure-density-operator}}
\begin{defisa}
	\puredensop & :: & \cpxmat \rightarrow \dB\\
	\puredensop\ \rho &\equiv & \exists \psi.\ \rho = \outerp{\psi}{\psi}
\end{defisa}
The density matrices that represent pure states are characterized as follows:
\begin{lemisa}
	\keyw{lemma}:\\
	\aindent \keyw{assumes}\ \rho \in \fcmats\ \keyw{and}\ \densop\ \rho\\
	\sindent \keyw{shows}\ \isalem{pure-density-charact}:\\
	\sindent\ (\puredensop\ \rho) \longleftrightarrow (\trc{\rho \cdot \rho} = 1)\\
	\sindent \keyw{and}\ \isalem{pure-density-charact'}:\\
	\sindent\ (\puredensop\ \rho) \longleftrightarrow (\rho \cdot \rho = \rho)
\end{lemisa}



One density operator that is used in the formalization of the measurement postulate is the density matrix that represents the so-called \emph{maximally mixed state}. It is the (unique) density matrix that is proportional to the identity matrix, and it is formalized as follows in Isabelle:
\newcommand{\maxmix}{\isacmd{max-mix-density}}
\begin{defisa}
	\maxmix & :: & \dN \rightarrow \cpxmat\\
	\maxmix\ n &= & \frac{1}{n}\cdot \idty_n
\end{defisa}
This is indeed a density matrix, and it admits the maximum von Neumann entropy, meaning that its spectrum admits the maximum Shannon entropy. In other words, this density matrix represents a physical system generating the maximum amount of information.	
\paragraph{The composite state postulate.} The composite state postulate is the following:
\begin{quote}
	\emph{The state space of a composite physical system is the tensor product of the state spaces of the component physical systems. If the system consists of $n$ individual systems and system $i$ is in the state represented by $\rho_i$ for $i\in \interv{1}{n}$, then the joint state of the composite system is in the state represented by $\rho_1\otimes \rho_2\otimes \cdots \otimes \rho_n$.}
\end{quote}
Consider two  physical systems $A$ and $B$, to which are associated the Hilbert spaces $\dH_A$ and $\dH_B$, and which are represented by the density matrices $\rho_A$ and $\rho_B$. Then the composite state postulate asserts that the state space of the composite system consisting of $A$ and $B$ is  $\dH_A\otimes \dH_B$,  and that the state of this system is represented by $\rho_A\otimes \rho_B$. All quantum systems to which are associated the Hilbert space $\dH_A\otimes \dH_B$ are said to be in a \emph{bipartite state}. Not all of these systems admit density matrices that can be written as convex combinations of  tensor products of the form $\rho\otimes \rho'$, hence the notions of \emph{separability} and \emph{entanglement}:
\begin{definition}
	A state represented by a density matrix $\rho$ is \emph{separable} if $\rho$ can be written as
	\[\rho\ =\ \sum_{i = 1}^n \lambda_i\cdot \rho^i_A\otimes \rho^i_B,\ \text{where for $i \in \interv{1}{n}$, $0 \leq \lambda_i\leq 1$, and $\sum_{i=1}^n \lambda_i = 1$.}\]
	Otherwise, the state is \emph{entangled}.
\end{definition}
Intuitively, if a system is in a separable state, then each of its components can be considered
independently from the other one, and this is not the case if the system is in an entangled state. Separable states are characterized by predicate {\sepdensity} in our formalization. 

\begin{example}\label{ex:compo}
	Let $\ket{\Psi_s} \isdef \ket{+}\otimes\ket{+}$, then\footnote{Lemma \isalem{outer-prod-tensor-comm}.} $\rho_s \isdef \outerp{\Psi_s}{\Psi_s} = (\outerp{+}{+})\otimes (\outerp{+}{+})$ is a density matrix that represents a 2-qubit system in a separable state.
\end{example}

\begin{example}\label{ex:bell}
	The so-called \emph{Bell states}, listed below along with their density matrices are all entangled states:
	\[\begin{array}{rclcrcl}
		\ket{\Phi^+} & \isdef & \frac{1}{\sqrt{2}}\left(\ket{00} + \ket{11}\right) & \text{and} & \rho_{\Phi^+} & \isdef & \outerp{\Phi^+}{\Phi^+}\\
		\ket{\Phi^-} & \isdef & \frac{1}{\sqrt{2}}\left(\ket{00} - \ket{11}\right) & \text{and} & \rho_{\Phi^-} & \isdef & \outerp{\Phi^-}{\Phi^-}\\
		\ket{\Psi^+} & \isdef & \frac{1}{\sqrt{2}}\left(\ket{01} + \ket{10}\right) & \text{and} & \rho_{\Psi^+} & \isdef & \outerp{\Psi^+}{\Psi^+}\\	
		\ket{\Psi^-} & \isdef & \frac{1}{\sqrt{2}}\left(\ket{01} - \ket{10}\right) & \text{and} & \rho_{\Psi^-} & \isdef & \outerp{\Psi^-}{\Psi^-}
	\end{array}\]
\end{example}
The fact that the components of a physical system cannot be considered independently from the other ones is counterintuitive, and led to the formulation of the famous EPR paradox that was presented in the Introduction.

\subsection{System evolution and measurement}

\paragraph{The evolution postulate.} The evolution postulate is the following:
\begin{quote}
	\emph{The evolution of a closed system is described by a unitary transformation: the density matrix $\rho$ representing the state of the system at time $t$ is related to the density matrix $\rho'$ representing the state of the system at time $t'$  by the equation $\rho' = U\cdot \rho\cdot \adj{U}$,  where $U\isdef U(t,t')$ is a unitary operator.}
\end{quote}
\newcommand{\hadam}{\mathbf{H}}
\newcommand{\cnot}{\textsc{cnot}}
\newcommand{\bellop}{\mathbf{B}}
Intuitively, a closed system is one that has no interaction with an external environment.
In Quantum Information theory, the unitary operators that are considered are constant over time, and can thus be represented by matrices with complex entries.

\begin{example}
	The \emph{Hadamard operator} is an operator that acts on a single qubit, it is represented by the matrix
	\[\hadam\ \isdef\ \frac{1}{\sqrt{2}}\cdot\begin{pmatrix}
		1 & 1\\
		1 & -1
	\end{pmatrix}.\]
	We have $\hadam\cdot\ket{0} = \frac{1}{\sqrt{2}}\begin{pmatrix}
		1\\ 1
	\end{pmatrix} = \ket{+}$, hence if $\rho_0 = \outerp{0}{0}$ then
	\[\hadam\cdot\rho_0\cdot\adj{\hadam}\ =\ \hadam\cdot\outerp{0}{0} \cdot \adj{\hadam}\ =\ \left(\hadam\cdot\ket{0}\right)\cdot \adj{\left(\hadam\cdot\ket{0}\right)}\ =\ \outerp{+}{+}.\] 
	The identity matrix is clearly a unitary operator, hence so is $\hadam\otimes \idty$ (Lemma \isalem{tensor-mat-unitary}). This is an operator that acts on two qubits, and transforms the first qubit according to the Hadamard operator while leaving the second one unchanged.
\end{example}

\begin{example}
	The \emph{Controlled-not operator} is an operator that acts on two qubits; intuitively, it performs a \textsc{not} operation on the second qubit exactly when the first one is $\ket{1}$. It is represented by the matrix
	\[\cnot\ \isdef\ \begin{pmatrix}
		1 & 0 & 0 & 0\\
		0 & 1 & 0 & 0\\
		0 & 0 & 0 & 1\\
		0 & 0 & 1 & 0
	\end{pmatrix}.\]
	Operators can of course be composed. For example, $\bellop \isdef \cnot\cdot (\hadam\otimes \idty)$ is an operator that acts on two qubits. We have:
	\begin{eqnarray*}
		\bellop\cdot\ket{00} & = & \cnot\cdot (\hadam\otimes \idty)\cdot (\ket{0}\otimes \ket{0})\\
		& = & \cnot \cdot \left(\left(\frac{1}{\sqrt{2}}\cdot (\ket{0} + \ket{1})\right) \otimes \ket{0}\right)\\
		& = & \frac{1}{\sqrt{2}}\cdot \cnot\cdot \left(\ket{00} + \ket{10}\right)\\
		& = & \frac{1}{\sqrt{2}}\cdot \left(\ket{00} + \ket{11}\right)
	\end{eqnarray*}
	Thus, by letting $\rho_0 \isdef \outerp{0}{0}$, we have $\bellop \cdot \rho_0\cdot\adj{\bellop} = \rho_{\Phi^+}$: operator $\bellop$ transforms $\rho_0$ into one of the Bell states from Example \ref{ex:bell}.
\end{example}

\begin{remark}
	Because all unitary transformations are reversible, the evolution postulate entails constraints that do not occur in the classical setting when using Quantum Information techniques to perform tasks or solve problems.
\end{remark}

\paragraph{The measurement postulate.} We present the so-called projective measurement postulate. This formulation is the one that was used by von Neumann in his axiomatic treatment of quantum mechanics, it is based on \emph{projection-valued measures} (PVMs). Note that there exists a more general measurement postulate, based on \emph{positive operator-valued measures} (POVMs). We chose to formalize the projective measurement postulate because it is sufficient to obtain the required results on the CHSH inequality and Tsirelson's upper-bound, and there is no loss of generality since Naimark's Dilation Theorem guarantees that any POVM can be obtained from a PVM on a larger state space \cite{Naimark}.	
\begin{quote}
	\emph{A projective measurement is described by an observable --any physical quantity that can be measured--, which is represented by  a Hermitian operator $M$ on the state space of the observed system. The measurement outcomes for an observable are the eigenvalues of the associated Hermitian operator. Given the spectral decomposition
		\[M\ =\ \sum_{a \in \spct{M}} a\cdot \prjct{a},\]
		where $\prjct{a}$ is the (orthogonal) projector onto the eigenspace of $M$ with eigenvalue $a$, the probability of obtaining outcome $a$ when measuring the system represented by the density operator $\rho$ is $\trc{\prjct{a}\cdot\rho}$, and the state of the system after the measurement is represented by
		\[\rho'\ \isdef\ \frac{\prjct{a}\cdot\rho\cdot \prjct{a}}{\trc{\prjct{a}\cdot\rho}}.\]}
\end{quote}	
It is standard to identify observables with their representatives as Hermitian operators, we will follow this convention in the remainder of the paper.

		%
	%

\newcommand{\diagelems}{\texttt{diag-elems}}
\newcommand{\diagitoel}{\texttt{diag-idx-to-el}}
\newcommand{\diageli}{\texttt{diag-elem-indices}}
\newcommand{\prjvec}{\texttt{project-vecs}}
\newcommand{\mkmo}{\isacmd{mk-meas-outcome}}
\newcommand{\eigvals}{\isacmd{eigvals}}
\newcommand{\makepm}{\isacmd{make-pm}}
\newcommand{\schur}{\texttt{unitary-schur-decomposition}}
\newcommand{\decard}{\texttt{dist-el-card}}
\newcommand{\measout}{\isacmd{measure-outcome}}
\newcommand{\prjmeas}{\isacmd{proj-measurement}}
\newcommand{\injon}{\isacmd{inj-on}}
\newcommand{\moval}[1]{#1^\mathrm{v}}
\newcommand{\moprj}[1]{#1^\mathrm{p}}
\newcommand{\measopr}{\isacmd{meas-outcome-prob}}

Projective measurements can be formalized in several ways, and we chose to stick with a formalization that is as close as possible to the one used in \cite{liu19} for the quantum programs they consider, for the sake of future reusability. We consider a measure outcome as a couple $(\lambda, \prjct{\lambda})$, where $\lambda$ represents the output of the measure and $\prjct{\lambda}$ is the associated projector, and we introduce a predicate that characterizes  projective measurements. The first parameter of this predicate represents the number of possible measure outcomes and the second parameter is the collection of measure outcomes. We require that the values of the measure outcomes are pairwise distinct, that the associated projectors have the correct dimensions and are orthogonal projectors, that sum to the identity. For the sake of readability, if $M_i = (\lambda, \prjct{\lambda})$ is a measure outcome, then we denote $\lambda$ by $\moval{M_i}$ and $\prjct{\lambda}$ by $\moprj{M_i}$.

%
\begin{defisa}
	\multicolumn{3}{l}{\keyw{type-synonym}\ \measout\ =\ \dR\times \cpxmat} \lmspace\\
	\prjmeas & :: & \dN \rightarrow (\dN \rightarrow \measout) \rightarrow \dB\lmspace\\
	\prjmeas\ n\ M\ & \Leftrightarrow & \injon\ (\lambda i.\ \moval{M_i})\ \ \interv{0}{n-1}\ \wedge\\
	& & \forall j < n.\, (\moprj{M_j} \in \fcmats \wedge \projector\ \moprj{M_j})\ \wedge\\
	& & \forall i,j < n.\, (i\neq j \Rightarrow \moprj{M_i}\cdot \moprj{M_j} = \vzero)\ \wedge\\
	& & \sum_{j = 0}^{n-1} \moprj{M_j} = \idty	
\end{defisa}
The projective measurement predicate states that, the probability of obtaining result $\lambda$ when measuring the density operator $\rho$ is $\trc{\rho \prjct{\lambda}}$. Although $\rho$ and $\prjct{\lambda}$ are complex matrices, these traces are real positive numbers that sum to 1.
\begin{defisa}
	\measopr & :: & \cpxmat \rightarrow (\dN \rightarrow \measout) \rightarrow \\
	& & \quad \quad \dN \rightarrow \dC\\
	\measopr\ \rho\ M\ i & = & \trc{\rho \cdot \moprj{M_i}}
\end{defisa}

\isatext{\[\begin{array}{l}
		\keyw{lemma}\ \isalem{meas-outcome-prob-real}:\\
		\aindent \keyw{assumes}\ \rho \in \fcmats\ \keyw{and}\ \densop\ \rho\\
		\aindent \keyw{and}\ \prjmeas\ n\ M\ \keyw{and}\ i < n\\
		\sindent \keyw{shows}\ \measopr\ \rho\ M\ i \in \dR\lmspace\\
		\keyw{lemma}\ \isalem{meas-outcome-prob-pos}:\\
		\aindent \keyw{assumes}\ \rho \in \fcmats\ \keyw{and}\ \densop\ \rho\\
		\aindent \keyw{and}\ \prjmeas\ n\ M\ \keyw{and}\ i < n\\
		\sindent \keyw{shows}\ \measopr\ \rho\ M\ i \geq 0\lmspace\\
		\keyw{lemma}\ \isalem{meas-outcome-prob-sum}:\\
		\aindent \keyw{assumes}\ \rho \in \fcmats\ \keyw{and}\ \densop\ \rho\\
		\aindent \keyw{and}\ \prjmeas\ n\ M\\
		\sindent \keyw{shows}\ \sum_{j = 1}^{n-1} (\measopr\ \rho\ M\ j) = 1
	\end{array}\]}

When the outcome of the projective measurement of $\rho$ is $\lambda$, $\rho$ collapses into $\frac{\prjct{\lambda}\rho \prjct{\lambda}}{\trc{\rho \prjct{\lambda}}}$. When formalizing this collapse in Isabelle, some care must be taken to handle the case of results that occur with probability zero. Although such cases are never meant to be considered when analyzing the result of a measurement, it is still necessary to provide a reasonable definition of the state $\rho$ collapses into. We have chosen to make $\rho$ collapse into the maximally mixed state in this case, so that after a measurement, a density matrix always collapses into a density matrix (Lemma \isalem{density-collapse-operator}).

\newcommand{\dtycol}{\isacmd{density-collapse}}

\isatext{\[\begin{array}{lcl}
		\dtycol & :: & \cpxmat \rightarrow \cpxmat \rightarrow \cpxmat\\
		\dtycol\ \rho\ \Pi & = & \keyw{if}\ \trc{\rho \cdot \Pi} = 0\\
		& & \keyw{then}\ \maxmix\ (\dimrow\ \rho)\\
		& & \keyw{else}\ \frac{\Pi\cdot\rho\cdot \Pi}{\trc{\rho\cdot \Pi}}
	\end{array}
	\]}


\begin{example}\label{ex:meas:1}
	Consider $A \isdef 	\begin{pmatrix}
		1 & 0\\
		0 & -1
	\end{pmatrix}$. This is an observable with eigenvalues $\pm 1$, its decomposition according to its projectors is $A = \outerp{0}{0} - \outerp{1}{1}$. 
	\begin{itemize}
		\item The measurement by $A$ of the state represented by the density matrix $\outerp{0}{0}$ produces the outcome $1$ with probability
		\[\trc{(\outerp{0}{0})\cdot (\outerp{0}{0})} = \trc{\outerp{0}{0}} = 1,\]
		and the state after the measurement is represented by 
		\[(\outerp{0}{0})\cdot (\outerp{0}{0}) \cdot (\outerp{0}{0}) = \outerp{0}{0}.\]
		\item The measurement by $A$ of the state represented by the density matrix $\outerp{+}{+}$ produces outcome $1$ with probability
		\begin{align*}
			\trc{(\outerp{0}{0})\cdot (\outerp{+}{+})} &= \frac{1}{2}\cdot \trc{(\outerp{0}{0})\cdot \big(\outerp{0}{0} +\outerp{0}{1} +\outerp{1}{0} + \outerp{1}{1}\big)}\\
			& = \frac{1}{2}\cdot \trc{\outerp{0}{0} + \outerp{0}{1}}\\
			& = \frac{1}{2}\cdot \left(\trc{\outerp{0}{0}} + \trc{\outerp{0}{1}}\right)\\
			&= \frac{1}{2},
		\end{align*}
		and the state after the measurement is represented by 
		\begin{align*}
			\rho' &= 2\cdot (\outerp{0}{0})\cdot (\outerp{+}{+}) \cdot (\outerp{0}{0})\\
			& = 2\cdot \frac{1}{2} \cdot (\outerp{0}{0})\cdot \big(\outerp{0}{0} +\outerp{0}{1} +\outerp{1}{0} + \outerp{1}{1}\big) \cdot (\outerp{0}{0})\\
			& = \left(\outerp{0}{0} + \outerp{0}{1}\right) \cdot (\outerp{0}{0})\\
			& = \outerp{0}{0}.
		\end{align*}
		Similarly, the measurement produces the outcome $-1$ with probability $\frac{1}{2}$ and the state after the measurement is represented by $\outerp{1}{1}$.
	\end{itemize}
\end{example}
\begin{remark}
	Observable $A$ from Example \ref{ex:meas:1} represents measurements of a single qubit in the standard basis, a notion that was already defined in \cite{isa-dirac}. Our setting permits to represent measurements in arbitrary bases, as evidenced, e.g., by the observable $A'\isdef (\outerp{+}{+}) - (\outerp{-}{-})$.
\end{remark}

\begin{example}
	Consider the density matrices $\rho_s$ and $\rho_{\Phi^+}$ from Examples \ref{ex:compo} and \ref{ex:bell} respectively, and matrix $A$ from Example \ref{ex:meas:1}. The matrices 
	\[A_l\isdef A\otimes \idty_2 = (\outerp{0}{0}\otimes \idty_2) - (\outerp{1}{1}\otimes \idty_2) \quad \text{and}\quad A_r \isdef \idty_2\otimes A\] 
	are both observables (Lemma \isalem{tensor-mat-hermitian}), which can be viewed as \emph{local} measurements of the first (for $A_l$) and second qubit (for $A_r$) of a system. We have the following:
	\begin{itemize}
		\item A measurement of $\rho_s$ by $A_l$ yields outcome $1$ with probability $\frac{1}{2}$, and the resulting state is $(\outerp{0}{0})\otimes (\outerp{+}{+})$. A measurement of this state by $A_r$ yields outcome $1$ with probability $\frac{1}{2}$ and outcome $-1$ with probability $\frac{1}{2}$. A measurement of $\rho_s$ by $A_l$ yields outcome $-1$ with probability $\frac{1}{2}$, and a measurement of the resulting state by $A_r$ yields outcomes $1$ and $-1$, both with probability $\frac{1}{2}$.
		\item A measurement of $\rho_{\Phi^+}$ by $A_l$ yields outcome $1$ with probability $\frac{1}{2}$, and the resulting state is $(\outerp{0}{0})\otimes (\outerp{0}{0})$. A measurement of this state by $A_r$ yields outcome $1$ with probability $1$. Similarly, a measurement of $\rho_{\Phi^+}$ by $A_l$ yields outcome $-1$ with probability $\frac{1}{2}$, and a measurement of the resulting state by $A_r$ yields outcome $-1$ with probability $1$.
	\end{itemize}
	The fact that, once the outcome of the measurement of the first qubit of a (pure)
	Bell state is known, the outcome of the  measurement of the second qubit is also known with certainty is the basis of the EPR paradox \cite{EPR}, and appears to suggest that there was an instantaneous transmission of information from the first qubit to the second one, although they may be physically separated.
\end{example}

%

We briefly describe the construction of a projective measurement for a given observable ({\makepm} in our formalization). The construction relies on the fact that observables are represented by Hermitian matrices, and that a Hermitian matrix $A$ can be decomposed as $A = U\cdot B\cdot \adj{U}$, where $B$ is a diagonal matrix and $U$ is unitary. A  construction of $B$ and $U$ based on the Schur decomposition theorem is available in Isabelle; this theorem was developed in \cite{jordan} and extended in \cite{liu19}. The projective measurement for $A$ is constructed using the fact that the spectrum of $A$ consists of the diagonal elements of $B$, and because $U$ is unitary, its column vectors are necessarily normalized and pairwise orthogonal. 
	When $A$ is an observable, $\makepm\ A$ is indeed a projective measurement and the original matrix can be recovered by summing the projectors scaled by the corresponding eigenvalues:
	\isatext{\[\begin{array}{l}
			\keyw{lemma}\ \isalem{make-pm-proj-measurement}:\\
			\aindent \keyw{assumes}\ A \in \fcmats\ \keyw{and}\ \hermitian\ A\\
			\aindent \keyw{and}\ \makepm\ A = (n, M)\\
			\sindent \keyw{shows}\ \prjmeas\ n\ M\lmspace\\	
			\keyw{lemma}\ \isalem{make-pm-sum}:\\
			\aindent \keyw{assumes}\ A \in \fcmats\ \keyw{and}\ \hermitian\ A\\
			\aindent \keyw{and}\ \makepm\ A = (n, M)\\
			\sindent \keyw{shows}\ \sum_{i = 0}^{n-1} \moval{M_i}\cdot \moprj{M_i} = A
		\end{array}\]}
	
	\newcommand{\expectval}{\isacmd{expect-value}}
	Assume it is possible to prepare several systems in the same state, i.e., represented by the same density matrix, and that projective measurements are performed on all the systems. The average value of the measurement outcomes is called the \emph{expectation value} of the system:
	\begin{defisa}
		\expectval & :: & \cpxmat \rightarrow \dN \times (\dN \rightarrow \measout) \rightarrow \dC\\
		\expectval\ \rho\ n\ M & = & \sum_{i=0}^{n-1} (\measopr\ \rho\ M\ i)\cdot \moval{M_i}
	\end{defisa}
		\newcommand{\exval}[2]{\tuple{#1}_{#2}}
		Given an observable $A$, the expectation value of a system represented by the density matrix $\rho$ for $A$ is denoted by $\exval{A}{\rho}$. This value can be computed in a straightforward way:
		\begin{lemisa}
			\keyw{lemma}\ \isalem{expect-value-hermitian}:\\
			\aindent \keyw{assumes}\ \rho\in \fcmats\ \keyw{and}\ A \in \fcmats\ \keyw{and}\ \hermitian\ A\\
			\aindent \keyw{and}\ \makepm\ A = (n, M)\\
			\sindent \keyw{shows}\ \expectval\ \rho\ n\ M\ = \trc{A\cdot\rho}
		\end{lemisa}
		
		Consider a projective measurement $(m,M)$ corresponding to an observable $A$ on a Hilbert space $\dH$, and a projective measurement $(n,N)$ corresponding to an observable $B$ on a Hilbert space $\dH'$. It is possible to relate the projective measurement corresponding to $A\otimes B$ on the Hilbert space $\dH\otimes \dH'$ with the couple $(m\cdot n, \tilde{Q})$, where the elements of $\tilde{Q}$ are of the form $(\moval{M_i}\cdot \moval{N_j}, \moprj{M_i}\otimes \moprj{N_j})$ for $i < m$ and $j < n$. Note that this couple is not necessarily a projective measurement, because the injectivity of the outcomes is not guaranteed. Still, it induces a probability distribution on the couple of measure outcomes when measuring a system in state $\rho$ by setting the probability of obtaining outcome $(\moval{M_i}, \moval{N_j})$ to be $p_\rho (\moval{M_i}, \moval{N_j}) \isdef \trc{(\moprj{M_i}\otimes \moprj{N_j})\cdot\rho}$. 
		These probabilities, also referred to as \emph{correlations}, are positive and sum up to 1. The average value of the corresponding outcome product when measuring a system in state $\rho$ is derived as follows (Lemma \isalem{tensor-mat-make-pm-mult-trace} in our formalization):
		\begin{align*}
			\sum_{i<m, j<n} \moval{M_i}\cdot \moval{N_j}\cdot p_\rho (\moval{M_i}, \moval{N_j}) &= \sum_{i<m, j<n} \moval{M_i}\cdot \moval{N_j}\cdot \trc{(\moprj{M_i}\otimes \moprj{N_j})\cdot\rho}\\
			& = \trc{\left(\sum_{i<m}\moval{M_i}\cdot \moprj{M_i}\right)\otimes \left(\sum_{j<n}\moval{N_j}\cdot \moprj{N_j}\right)\cdot \rho}\\
			& = \trc{(A\otimes B)\rho}\\
			& = \tuple{A\otimes B}_\rho.
		\end{align*}
		
		\section{The CHSH inequality}\label{sec:chsh}
		\newcommand{\qprobs}[2]{\mathrm{p}_{#2}(#1)}
		\newcommand{\gscore}{\mathcal{C}}
		\newcommand{\score}[1]{\mathrm{S}_{#1}}
		
		We provide a high-level overview of the CHSH experiment.
		Consider a game involving two players, Alice and Bob, who can communicate before each round of the game to devise a strategy but are not allowed to communicate with one another once the round starts. At each round of the game, a third party randomly draws two requests  $x,y\in \set{0,1}$, sends request $x$ to Alice and $y$ to Bob. Note that Alice does not know the value of input $y$, and Bob does not know the value of input $x$. Alice outputs an answer $a\in \set{-1,1}$, and Bob an answer $b\in \set{-1,1}$; these answers are used to compute their score at each round. The score is defined by the following rule: the players score  $1$ point  when $x=y=0$  and they provide distinct outputs, or when at least one of $x,y$ is $1$ and they provide the same output. Otherwise, they score $-1$ point. At the end of the experiment, 
		the sequence of outcomes can be used to estimate the probabilities $\qprobs{a_x,b_y}{x,y}$ that Alice's measure outcome is $a_x\in \set{-1,1}$ and that Bob's is $b_y\in\set{-1,1}$ when Alice was given input $x$ and Bob was given input $y$. In turn,  these probabilities permit to compute the 
		maximum  number of points that Alice and Bob can score on average, depending on their inputs. 
		If $\score{x,y}$ denotes this  score 
		for inputs $x$ and $y$, then we have the following equalities:
		\begin{itemize}
			\item When $(x,y) \neq (0,0)$:
			\begin{align*}
				\score{x,y} & = \qprobs{-1, -1}{x,y} + \qprobs{1, 1}{x,y} - \qprobs{1, -1}{x,y} - \qprobs{-1, 1}{x,y}\\
				& = \sum_{a_x,b_y \in \set{-1,1}} a_x\cdot b_y\cdot \qprobs{a_x, b_y}{x,y}\\ 
				& = \dE_{x,y}[a_x\cdot b_y].
			\end{align*}
			\item When $x=0$ and $y=0$:
			\begin{align*}
				\score{0,0} & = \qprobs{1, -1}{0,0} + \qprobs{-1, 1}{0,0} - \qprobs{-1, -1}{0,0} - \qprobs{1, 1}{0,0}\\
				& = -\dE_{0,0}[a_0,b_0].
			\end{align*}
		\end{itemize}
		Let $\gscore \isdef \dE_{0,1}[a_0\cdot b_1] - \dE_{0,0}[a_0,b_0] + \dE_{1,0}[a_1\cdot b_0] + \dE_{1,1}[a_1\cdot b_1]$.
		Using the fact that $x$ and $y$ are drawn uniformly at random, the average score  of Alice and Bob is therefore $\frac{1}{4}\cdot \gscore$.
		It can be shown that when the players can only act locally using classical resources, regardless of the strategy devised by Alice and Bob which can be deterministic or probabilistic, it holds that $\gscore \leq 2$. See \cite{scarani2019bell} for details.
		
		However, Alice and Bob can devise new strategies to play the CHSH game when they have access to quantum resources.
		In this setting, Alice and Bob can perform measurements of two observables -- $A_0$ and $A_1$ for Alice, $B_0$ and $B_1$ for Bob --, all with eigenvalues $\pm 1$, and at each round they both possess part of a quantum system in a given bipartite state. One such strategy is the following.
		When sent inputs $x$ and $y$, Alice and Bob respectively select observables $A_x$ and $B_y$ and simultaneously\footnote{As mentioned in \cite{nielsen-book}, it would be more precise to write \emph{in a causally disconnected manner}, meaning that no information can be transmitted between Alice and Bob. This is  ensured in practice by physically separating them, assuming that  information cannot travel  faster than light.} measure these observables on the part of the system they receive. They output the (classical) result of their measurement, and their average score is computed similarly to the classical case. This score admits the same upper-bound in the classical setting and under the local hidden variable hypothesis as formulated by Einstein, but it is violated in the  quantum setting, as experimentally verified by Aspect.

		\newcommand{\chshop}{\isacmd{CHSH-op}}
		\newcommand{\conjfield}{\isacmd{conjugatable-field}}
		\newcommand{\chshhermit}{\isacmd{CHSH-cond-hermit}}
		\newcommand{\chshexp}{\isacmd{CHSH-expect}}
		\newcommand{\chshlocal}{\isacmd{CHSH-cond-local}}

				\subsection{The local hidden-variable assumption and a counterexample}
				\newcommand{\integrable}{\isacmd{integrable}}
				Intuitively, a hidden-variable theory can be viewed as an attempt to restore determinism to quantum mechanics, by introducing objects that, if observed, would make measurements deterministic. Local hidden-variable theories are designed to further ensure \emph{quantum locality}, i.e., prevent the distant effects of a density collapse after a measurement, as these effects appear to be caused by faster-than-light information transmission. More formally, according to the local hidden-variable assumption on a bipartite quantum system, there exists a probability space and independent random variables such that, when performing simultaneous measurements on the system, the outcome probabilities (the values of which are given by the measurement postulate) are expectations in the underlying probability space. 
				This hypothesis is often found in articles and textbooks under the assumption that the probability space admits a density\footnote{Including in the original paper on the CHSH inequality \cite{CHSH}.} in the measure-theoretic sense, but it is defined in our formalization in a more general case, where no assumption on the existence of a density is made, and properties on the considered random variables are assumed to hold almost everywhere rather than on the entire probability space. 
				
				We formalize the local hidden-variable hypothesis by considering random variables that are positive almost everywhere ({\posrv} below), and collections of random variables that sum up to 1 almost everywhere ({\psrv} below). Given a bipartite state represented by the density matrix $\rho$ and two observables $A$ and $B$ representing the local measurements that are carried out, the local hidden-variable hypothesis states that it is possible to associate to $\rho, A, B$ a probability space $\calM$ and collections of random variables $X$ and $Y$, where each $X_a\in X$ is associated with an eigenvalue $a$ of $A$ (and thus to a measurement outcome) and similarly for each random variable $Y_b \in Y$ and the eigenvalues of $B$. These random variables are related to the probabilities of obtaining the measurement outcome $(a,b)$ by the equation $p_{A,B}^\rho(a,b) = \trc{\prjct{a}\cdot \prjct{b}\cdot \rho} = \dE[X_a\cdot Y_b]$.
				\isatext{\[\begin{array}{lcl}
						\posrv & :: & \alpha\ \msr \rightarrow (\alpha\rightarrow \dR) \rightarrow \dB\\
						\posrv\ \calM\ X & \equiv & X\in \bmeas\ \calM\ \wedge\ \AEv_\calM\, x.\  X(x) \geq 0\lmspace\\
						\psrv & :: & \alpha\ \msr \rightarrow \cpxmat \rightarrow (\dC\rightarrow \alpha \rightarrow \dR) \rightarrow \dB\\
						\psrv\ \calM\ A\ X & \equiv & \AEv_\calM\, x. \sum_{a\in \spct{A}} X_a(x) = 1\lmspace\\
						\lhv & :: & \alpha\ \msr \rightarrow \cpxmat \rightarrow\\
						& & \quad \cpxmat \rightarrow \cpxmat \rightarrow\\
						& & \quad  (\dC \rightarrow \alpha \rightarrow \dR) \rightarrow (\dC \rightarrow \alpha \rightarrow \dR) \rightarrow \dB\\
						\lhv\ \calM\ A\ B\ \rho\ X\ Y & \equiv & \pbspace\ \calM\ \wedge\\
						& & \psrv\ \calM\ A\ X\ \wedge\ \psrv\ \calM\ B\ Y\ \wedge\\
						& & \forall a\in \spct{A}.\, \posrv\ \calM\ X_a\ \wedge\\
						& & \forall b\in \spct{B}.\, \posrv\ \calM\ Y_b\ \wedge\\
						& & \forall a\in \spct{A}.\, \forall b\in \spct{B}.\\ 
						& & \quad \Bigl(\integrable\ \calM\ (X_a\cdot Y_b)\ \wedge\\
						& & \quad \ \ \dE[X_a\cdot Y_b] = \trc{\prjct{a}\cdot \prjct{b}\cdot \rho}\Bigr)
					\end{array}\]}
				
				In the CHSH game, after Alice and Bob have been provided requests $x$ and $y$ and measured observables $A_x$ and $B_y$ on their part of the quantum system $\rho$ several times, the probabilities of their outcomes are given by
				\[\qprobs{a_x,b_y}{x,y} = p_{A_x,B_y}^\rho(a_x,b_y) =  \dE[X_{a_x}\cdot Y_{b_y}],\]
				and the corresponding expectations used to compute the average score are
				\begin{align*}
					\dE_{x,y}[a_x\cdot b_y] & = \sum_{a_x,b_y\in \set{-1,1}} a_x\cdot b_y\cdot p_{A_x,B_y}^\rho(a_x,b_y)\\
					& = \sum_{a_x,b_y\in \set{-1,1}} a_x\cdot b_y\cdot\dE[X_{a_x}\cdot Y_{b_y}]\\
					& = \dE\left[\left(\sum_{a_x \in \set{-1,1}}a_x\cdot X_{a_x}\right)\cdot \left(\sum_{b_y \in \set{-1,1}}b_y\cdot Y_{b_y}\right)\right].
				\end{align*}
				
				\newcommand{\qtexp}{\isacmd{qt-expect}}
				
				Given a random variable $X$ that represents an observable $A$, we define the following random variable related to the expectation value of $A$:
				\begin{defisa}
					\qtexp & :: & \cpxmat\rightarrow (\dC\rightarrow \alpha \rightarrow \dR) \rightarrow \alpha \rightarrow \dR\\
					\qtexp\ A\ X & = & \left(\lambda \omega.\, \sum_{a\in \spct{A}} a \cdot X_a(\omega)\right)
				\end{defisa}
				Under the local hidden-variable hypothesis, expectation $\dE_{x,y}[a_x\cdot b_y]$ is therefore exactly $\dE[(\qtexp\ A_x\ X)\cdot(\qtexp\ B_y\ Y)]$. 
				We also have the following equality relating the expectation of this product of  random variables with the expectation values for the corresponding observables:
				\begin{lemisa}
					\keyw{lemma}\ \isalem{sum-qt-expect}:\\
					\aindent \keyw{assumes}\ \lhv\ \calM\ A\ B\ \rho\ X\ Y\\
					\aindent \keyw{and}\ \hermitian\ A\ \keyw{and}\ \hermitian\ B\\	
					\sindent \keyw{shows}\ \dE[(\qtexp\ A\ X)\cdot(\qtexp\ B\ Y)] = \trc{A\cdot B\cdot \rho}
				\end{lemisa}
				
				In order to relate the average score of the CHSH game to quantum measurements, we begin by defining a CHSH operator as follows\footnote{Note that these notions are formalized in a general setting that makes no mention of bipartite states or tensor products. }:
				\begin{defisa}
					\chshop & :: & \beta\ \gmat \rightarrow \beta\ \gmat \rightarrow \beta\ \gmat \rightarrow \beta\ \gmat \rightarrow \beta\ \gmat\\
					\chshop\ A_0\ A_1\ B_0\ B_1 & = & A_0\cdot B_1 - A_0\cdot B_0 + A_1\cdot B_0 + A_1\cdot B_1
				\end{defisa}
				The key properties the matrices $A_0, A_1, B_0$ and $B_1$ are meant to satisfy are represented by predicate {\chshhermit}: for $i,j \in \set{0,1}$, $A_i$ and $B_j$ must be Hermitian matrices of appropriate dimensions such that $A_i^2 = B_j^2 = \idty$, and $A_i\cdot B_j= B_j\cdot A_i$. 
				When the predicate {\chshhermit} is satisfied, the operator {\chshop} is a Hermitian matrix that thus represents an observable (Lemma \isalem{CHSH-op-hermitian}). The expectation value of a system represented by the density matrix $\rho$ for this operator is given by {\chshexp}:
				\begin{defisa}
					\chshexp & :: & \beta\ \gmat \rightarrow \beta\ \gmat \rightarrow \beta\ \gmat \rightarrow \beta\ \gmat \rightarrow \beta\ \gmat \rightarrow \beta\\
					\chshexp\ A_0\ A_1\ B_0\ B_1\ \rho & = & \trc{(\chshop\ A_0\ A_1\ B_0\ B_1)\cdot \rho}
				\end{defisa}
				Under the local hidden-variable hypothesis, the value $\gscore$ that can be achieved by Alice and Bob both sharing parts of a quantum system represented by $\rho$ is therefore exactly \isatext{$\chshexp\ A_0\ A_1\ B_0\ B_1\ \rho$}.
				
				When all joint measurements can be represented by local hidden variables, the following upper-bound is derived on the expectation value of a system with density matrix $\rho$ for the CHSH operator (the fact that $\rho$ is actually a density matrix with trace $1$ is not necessary to derive the upper-bound):
				\begin{lemisa}
					\keyw{lemma}\ \isalem{CHSH-expect-lhv-leq}:\\
					\aindent \keyw{assumes}\ \isacmd{positive}\ \rho\\
					\aindent \keyw{and}\ \chshhermit\ n\ A_0\ A_1\ B_0\ B_1\\
					\aindent \keyw{and}\ \lhv\ \calM\ A_0\ B_1\ \rho\ U_0\ V_1\\
					\aindent \keyw{and}\ \lhv\ \calM\ A_0\ B_0\ \rho\ U_0\ V_0\\
					\aindent \keyw{and}\ \lhv\ \calM\ A_1\ B_0\ \rho\ U_1\ V_0\\
					\aindent \keyw{and}\ \lhv\ \calM\ A_1\ B_1\ \rho\ U_1\ V_1\\
					\sindent \keyw{shows}\ \card{\chshexp\ A_0\ A_1\ B_0\ B_1\ \rho} \leq 2	
				\end{lemisa}
				In other words, as in the classical setting, the upper-bound of $\gscore$ in the quantum setting with the local hidden-variable hypothesis is $2$.
				
				\subsubsection*{A counterexample}
				
				\newcommand{\tX}{\isacmd{X}}
				\newcommand{\tZ}{\isacmd{Z}}
				\newcommand{\tXpZ}{\isacmd{XpZ}}
				\newcommand{\tZmX}{\isacmd{ZmX}}
				\newcommand{\tXI}{\isacmd{X-I}}
				\newcommand{\tZI}{\isacmd{Z-I}}
				\newcommand{\tIXpZ}{\isacmd{I-XpZ}}
				\newcommand{\tIZmX}{\isacmd{I-ZmX}}
				
				A suitable choice for a density matrix and observables permits to deduce that the local hidden-variable assumption cannot hold because the inequality from Lemma \isalem{CHSH-expect-lhv-leq} is violated. A standard choice is presented below.
				We consider the density operator 
				$\rho_{\Psi^-}\ =\ \outerp{\Psi^-}{\Psi^-}$, {where} $\ket{\Psi^-}\ =\ \frac{1}{\sqrt{2}}\left(\ket{01} - \ket{10}\right)$ {is one of the Bell states from Example \ref{ex:bell}},
				and we consider bipartite measurements of this entangled state. These measurements  involve the following observables:
				\[\begin{array}{rclcrcl}
					\tZ & \isdef & \begin{pmatrix}
						1 & 0\\
						0 & -1 \end{pmatrix} & \quad & \tX & \isdef & \begin{pmatrix}
						0 & 1\\
						1 & 0
					\end{pmatrix}\lmspace\\
					\tXpZ & \isdef & -\frac{1}{\sqrt{2}}(\tX + \tZ) & & \tZmX & \isdef & \frac{1}{\sqrt{2}}(\tZ - \tX)
				\end{array}\]
				The top two are those used by Alice, the other two are used by Bob. The corresponding separated measurements are represented by the following tensor products in our formalism:
				\[\begin{array}{rclcrcl}
					\tZI & \isdef & \tZ\otimes \idty & \quad & \tXI & \isdef & \tX \otimes \idty\\
					\tIXpZ & \isdef & \idty \otimes \tXpZ &\quad & \tIZmX & \isdef & \idty \otimes \tZmX
				\end{array}\]
				These observables satisfy the predicate {\chshhermit} (Lemma \isalem{limit-CHSH-cond}) and, together with density matrix $\rho_{\Psi^-}$, they permit to obtain the following equality:
				
				\begin{lemisa}
					\keyw{lemma}\ \isalem{CHSH-expect-limit}:\\
					\sindent \keyw{shows}\ \card{\chshexp\ (\tZ\otimes \idty)\ (\tX \otimes \idty)\ (\idty \otimes \tZmX)\ (\idty \otimes \tXpZ)\ \rho_{\Psi^-}} = 2\cdot\sqrt{2}	
				\end{lemisa}
				The upper-bound from the local hidden-variable hypothesis was indeed violated in the quantum setting by Aspect \cite{aspect}, who measured the polarizations of simultaneously emitted photons in his experiment. All questions about potential flaws in Aspect's experiment were definitively put to rest by subsequent, so-called loophole-free experiments  (see, e.g., \cite{hensen2015loophole, loophole2023}), establishing once and for all the impossibility for local hidden-variable theories to account for all predictions from quantum mechanics. 

			\subsection{Other upper-bounds}
			\newcommand{\gchsh}{\mathbf{S}}
			
			Lemma \isalem{CHSH-expect-limit} shows that local hidden-variable theories are not a suitable framework for quantum mechanics. There are still some interesting questions to answer: are there contexts in which the inequality of Lemma \isalem{CHSH-expect-lhv-leq} holds, even without any local hidden-variable assumption? And is it possible to know the maximum violation that can be reached in a quantum setting? The answer to the latter question was provided by Tsirelson in \cite{Tsirelson}, in what follows we formalize this result along with two conditions under which the CHSH inequality holds.
			
			First, the density matrix $\rho_{\Psi^-}$ that is used in Lemma \isalem{CHSH-expect-limit} is entangled; it turns out that this is a necessary condition for the CHSH inequality to be violated. Indeed, if $A_i$ and $B_j$ (for $i,j\in \set{0,1}$) are Hermitian matrices whose squares are identity matrices (predicate {\chshlocal}) and $\rho$ is a separable density, then the CHSH inequality cannot be violated:
			\begin{lemisa}
				\keyw{lemma}\ \isalem{CHSH-expect-separable-leq}:\\
				\aindent \keyw{assumes}\ \chshlocal\ n\ m\ A_0\ A_1\ B_0\ B_1\\
				\aindent \keyw{and}\ \sepdensity\ \rho\\		
				\sindent \keyw{shows}\ \card{\chshexp\ (A_0\otimes\idty_m)\ (A_1\otimes\idty_m)\ (\idty_n\otimes B_0)\ (\idty_n\otimes B_1)\ \rho_{\Psi^-}} \leq 2	
			\end{lemisa}
			In the general case, Tsirelson's upper-bound is obtained by noting that it is possible to bound the expectation value of a density matrix for an observable by the $\ltwo$ operator norm of the observable:
			\begin{lemisa}
				\keyw{lemma}\ \isalem{expect-val-L2-op-nrm}:\\
				\aindent \keyw{assumes}\ \densop\ \rho\\
				\sindent \keyw{shows}\ \card{\trc{A\cdot\rho}} \leq \nrm{A}
			\end{lemisa}
			An upper-bound for the $\ltwo$ operator norm of the CHSH operator is derived following \cite{scarani2019bell}. Fix the observables $A_0, A_1, B_0$ and $B_1$ such that the condition $\chshhermit\ n\ A_0\ A_1\ B_0\ B_1$ holds, and set $\gchsh \isdef \isatext{\chshop\ A_0\ A_1\ B_0\ B_1}$. Since $\gchsh$ is a Hermitian matrix, we have $\nrm{\gchsh} = \sqrt{\nrm{\gchsh^2}}$ (Lemma \isalem{hermitian-L2-op-nrm-sqrt}). Now, $\gchsh^2$ can be simplified as follows:
			\[\gchsh^2 = 4\cdot\idty_n - \commut{A_0}{A_1}\cdot\commut{B_0}{B_1}\quad \text{(Lemma \isalem{CHSH-op-square})},\]
			where $\commut{A}{B}\isdef A\cdot B - B\cdot A$ is called the \emph{commutator of $A$ and $B$}.
			The expression above permits to derive another upper-bound when one pair of observables -- $A_0$ and $A_1$ or $B_0$ and $B_1$ -- commutes. Indeed, in this case, one of $\commut{A_0}{A_1}$ or $\commut{B_0}{B_1}$ is the zero matrix and $\gchsh^2 = 4\cdot\idty_n$. This leads to the following lemma stating that the CHSH inequality cannot be violated when one pair of observables commutes:
			\begin{lemisa}
				\keyw{lemma}\ \isalem{CHSH-expect-commute-leq}:\\
				\aindent \keyw{assumes}\ \chshhermit\ n\ A_0\ A_1\ B_0\ B_1\\
				\aindent\keyw{and}\ \densop\ \rho\\
				\aindent \keyw{and}\ (\commut{A_0}{A_1} = \vzero_{n,n}) \vee (\commut{B_0}{B_1} = \vzero_{n,n})\\
				\sindent \keyw{shows}\ \card{\chshexp\ A_0\ A_1\ B_0\ B_1\ \rho} \leq 2
			\end{lemisa}
			When no pair of observables commutes, an upper-bound of $\nrm{\gchsh^2}$ is derived as follows:
			\[\begin{array}{rcll}
				\nrm{\gchsh^2} & \leq & 4\cdot \nrm{\idty_n} + \nrm{\commut{A_0}{A_1}\cdot\commut{B_0}{B_1}} & \text{(\isalem{L2-op-nrm-triangle})}\\
				& \leq & 4 + \nrm{\commut{A_0}{A_1}}\cdot\nrm{\commut{B_0}{B_1}} & \text{(\isalem{L2-op-nrm-mult-le})}\\
				& \leq & 4 + 4\cdot\nrm{A_0}\cdot\nrm{A_1}\cdot\nrm{B_0}\cdot\nrm{B_1} & \text{(\isalem{comm-L2-op-nrm-le})}\\
				& = & 8 & \text{(\isalem{herm-sq-id-L2-op-nrm})}
			\end{array}\]
			This permits to obtain Tsirelson's general upper-bound:
			\begin{lemisa}
				\keyw{lemma}\ \isalem{CHSH-expect-gen-leq}:\\
				\aindent \keyw{assumes}\ \chshhermit\ n\ A_0\ A_1\ B_0\ B_1\\
				\aindent\keyw{and}\ \densop\ \rho\\
				\sindent \keyw{shows}\ \card{\chshexp\ A_0\ A_1\ B_0\ B_1\ \rho} \leq 2\cdot\sqrt{2}
			\end{lemisa}
			Lemma \isalem{CHSH-expect-limit} shows that this bound is tight.
			
			\section{Conclusion}\label{sec:conclusion}
			In this paper we have shown how fundamental notions of quantum information and quantum computing can be formalized in Isabelle/HOL. The postulates of quantum mechanics were formalized using the general notion of density matrices which permit to represent mixed quantum states in a succinct way. Although the measurement postulate was presented with projection-valued measures instead of the more general positive operator-valued measures, this is with no loss of generality and we are in the process of formalizing Naimark's dilation theorem that shows that any positive operator-valued measure can be viewed as a projection-valued measure in a higher-dimensional Hilbert space.
			The formalization of results related to the CHSH inequality show that Isabelle/HOL can be used to certify fundamental results from quantum information, and we are exploring some of the consequences of the (non-)violation of this inequality.
			
			Lemma \isalem{CHSH-expect-commute-leq} states that the CHSH inequality cannot be violated when either Alice or Bob uses a pair of observables that commute. Commuting observables play an important role in quantum physics, where they represent quantities that can be measured simultaneously with an arbitrary precision, and are thus closely related to Heisenberg's uncertainty principle. From a mathematical point of view, the measurement properties are a consequence of the fact that commuting observables can be diagonalized in a common basis. We formalized the generalization of this result in \cite{Commuting_Hermitian-AFP}: any nonempty set of pairwise commuting observables can be diagonalized in a common basis. This is a fundamental building block for the study of \emph{Complete Sets of Commuting Observables}, which can be used to construct a basis of a Hilbert space made only of eigenvectors that are common to all observables.
			
			Another line of research we are currently investigating is related to the maximum violation of the CHSH inequality. It can be proved that Tsirelson's upper-bound is reached exactly when the quantum state shared by Alice and Bob is maximally entangled. The topic of certifying which states were used in an experiment and what measurements were performed by estimating the probabilities of measurement outcomes, without any assumption on the physical apparatus that was used (\emph{device independence}) is a major topic in quantum information. We are currently working on the proof of maximal entanglement when Tsirelson's bond is reached. We will carry on by working on device-independent statements that will permit, e.g., users of quantum cryptographic protocols to have a certification of the safety of a protocol even if they do not trust the physical devices used in the experiment.

			\paragraph{Acknowledgments.} This work benefited from the funding program \emph{``Plan France 2030''} (ANR-22-PETQ-0007)  of
			the French National Research Agency.

	\bibliography{biblio}

\begin{thebibliography}{10}

\bibitem{aspect}
A.~Aspect.
\newblock Experimental tests of {B}ell's inequalities in atomic physics.
\newblock In I.~Lindgren, A.~Ros{\'e}n, and S.~Svanberg, editors, {\em Atomic
  Physics 8}, pages 103--128, Boston, MA, 1983. Springer US.

\bibitem{Ballarin14}
C.~Ballarin.
\newblock Locales: {A} module system for mathematical theories.
\newblock {\em J. Autom. Reasoning}, 52(2):123--153, 2014.

\bibitem{Bell64}
J.~S. Bell.
\newblock {On the Einstein Podolsky Rosen paradox}.
\newblock {\em Physics}, 1(3):195--200, 1964.

\bibitem{berta2022gap}
M.~Berta, F.~G. Brand{\~a}o, G.~Gour, L.~Lami, M.~B. Plenio, B.~Regula, and
  M.~Tomamichel.
\newblock On a gap in the proof of the generalised quantum {S}tein's lemma and
  its consequences for the reversibility of quantum resources.
\newblock {\em arXiv preprint arXiv:2205.02813}, 2022.

\bibitem{Boender_2015}
J.~Boender, F.~Kamm\"uller, and R.~Nagarajan.
\newblock Formalization of quantum protocols using {C}oq.
\newblock {\em Electronic Proceedings in Theoretical Computer Science},
  195:71--83, Nov 2015.

\bibitem{afp-isa-dirac}
A.~Bordg, H.~Lachnitt, and Y.~He.
\newblock Isabelle marries {D}irac: a library for quantum computation and
  quantum information.
\newblock {\em Archive of Formal Proofs}, November 2020.
\newblock \url{https://isa-afp.org/entries/Isabelle_Marries_Dirac.html}, Formal
  proof development.

\bibitem{isa-dirac}
A.~Bordg, H.~Lachnitt, and Y.~He.
\newblock Certified quantum computation in {I}sabelle/{HOL}.
\newblock {\em J. Autom. Reason.}, 65(5):691--709, 2021.

\bibitem{Bordg22}
A.~Bordg, L.~C. Paulson, and W.~Li.
\newblock Simple type theory is not too simple: Grothendieck's schemes without
  dependent types.
\newblock {\em Exp. Math.}, 31(2):364--382, 2022.

\bibitem{brandao2015reversible}
F.~G. Brandao and G.~Gour.
\newblock Reversible framework for quantum resource theories.
\newblock {\em Physical review letters}, 115(7):070503, 2015.

\bibitem{brandao2008entanglement}
F.~G. Brandao and M.~B. Plenio.
\newblock Entanglement theory and the second law of thermodynamics.
\newblock {\em Nature Physics}, 4(11):873--877, 2008.

\bibitem{brandao2010generalization}
F.~G. Brandao and M.~B. Plenio.
\newblock A generalization of quantum stein's lemma.
\newblock {\em Communications in Mathematical Physics}, 295:791--828, 2010.

\bibitem{brandao2010reversible}
F.~G. Brandao and M.~B. Plenio.
\newblock A reversible theory of entanglement and its relation to the second
  law.
\newblock {\em Communications in Mathematical Physics}, 295:829--851, 2010.

\bibitem{CharetonBBPV21}
C.~Chareton, S.~Bardin, F.~Bobot, V.~Perrelle, and B.~Valiron.
\newblock An automated deductive verification framework for circuit-building
  quantum programs.
\newblock In N.~Yoshida, editor, {\em Programming Languages and Systems - 30th
  European Symposium on Programming, {ESOP} 2021, Held as Part of the European
  Joint Conferences on Theory and Practice of Software, {ETAPS} 2021,
  Luxembourg City, Luxembourg, March 27 - April 1, 2021, Proceedings}, volume
  12648 of {\em Lecture Notes in Computer Science}, pages 148--177, Luxembourg,
  2021. Springer.

\bibitem{Tsirelson}
B.~S. Cirel'son.
\newblock Quantum generalizations of {B}ell's inequality.
\newblock {\em Lett Math Phys}, 4(2):93--100, 1980.

\bibitem{CHSH}
J.~F. {Clauser}, M.~A. {Horne}, A.~{Shimony}, and R.~A. {Holt}.
\newblock {Proposed Experiment to Test Local Hidden-Variable Theories}.
\newblock {\em Phys. {R}ev. {L}ett.}, 23(15):880--884, Oct. 1969.

\bibitem{Dalton20}
B.~J. Dalton, B.~M. Garraway, and M.~D. Reid.
\newblock Tests for einstein-podolsky-rosen steering in two-mode systems of
  identical massive bosons.
\newblock {\em Phys. Rev. A}, 101:012117, Jan 2020.

\bibitem{Durrett}
R.~Durrett.
\newblock {\em Probability : theory and examples}.
\newblock The Wadsworth \& Brooks/Cole statistics/probability series. Wadsworth
  Inc. Duxbury Press, Belmont, California, 1991.

\bibitem{ProjectiveAFP}
M.~Echenim.
\newblock Quantum projective measurements and the {CHSH} inequality.
\newblock {\em Archive of Formal Proofs}, March 2021.
\newblock \url{https://isa-afp.org/entries/Projective_Measurements.html},
  Formal proof development.

\bibitem{Commuting_Hermitian-AFP}
M.~Echenim.
\newblock Simultaneous diagonalization of pairwise commuting {H}ermitian
  matrices.
\newblock {\em Archive of Formal Proofs}, July 2022.
\newblock \url{https://isa-afp.org/entries/Commuting_Hermitian.html}, Formal
  proof development.

\bibitem{TsirelsonAFP}
M.~Echenim, M.~Mhalla, and C.~Mori.
\newblock The {CHSH} inequality: {T}sirelson's upper-bound and other results.
\newblock {\em Archive of Formal Proofs}, April 2023.
\newblock \url{https://isa-afp.org/entries/TsirelsonBound.html}, Formal proof
  development.

\bibitem{EPR}
A.~Einstein, B.~Podolsky, and N.~Rosen.
\newblock Can quantum-mechanical description of physical reality be considered
  complete?
\newblock {\em Phys. Rev.}, 47(10):777--780, May 1935.

\bibitem{linalg}
P.~A. Fuhrmann.
\newblock Linear systems and operators in {H}ilbert space.
\newblock {\em Proceedings of the Edinburgh Mathematical Society},
  26(1):113--114, 1983.

\bibitem{Naimark}
I.~Gelfand and M.~Naimark.
\newblock On the imbedding of normed rings into the ring of operators in
  hilbert space.
\newblock {\em Matematiceskij sbornik}, 54(2):197--217, 1943.

\bibitem{hensen2015loophole}
B.~Hensen, H.~Bernien, A.~E. Dr{\'e}au, A.~Reiserer, N.~Kalb, M.~S. Blok,
  J.~Ruitenberg, R.~F. Vermeulen, R.~N. Schouten, C.~Abell{\'a}n, et~al.
\newblock Loophole-free {B}ell inequality violation using electron spins
  separated by 1.3 kilometres.
\newblock {\em Nature}, 526(7575):682--686, 2015.

\bibitem{hoelzl2012thesis}
J.~H{\"o}lzl.
\newblock {\em Construction and Stochastic Applications of Measure Spaces in
  Higher-Order Logic}.
\newblock PhD thesis, Institut f{\"u}r Informatik, Technische Universit{\"a}t
  M{\"u}nchen, October 2012.

\bibitem{ji2021mip}
Z.~Ji, A.~Natarajan, T.~Vidick, J.~Wright, and H.~Yuen.
\newblock {MIP}*= {RE}.
\newblock {\em Communications of the ACM}, 64(11):131--138, 2021.

\bibitem{Kammuller19}
F.~Kamm{\"{u}}ller.
\newblock Attack trees in {I}sabelle extended with probabilities for quantum
  cryptography.
\newblock {\em Comput. Secur.}, 87, 2019.

\bibitem{types-sets}
O.~Kuncar and A.~Popescu.
\newblock From types to sets by local type definition in higher-order logic.
\newblock {\em J. Autom. Reason.}, 62(2):237--260, 2019.

\bibitem{liu19}
J.~Liu, B.~Zhan, S.~Wang, S.~Ying, T.~Liu, Y.~Li, M.~Ying, and N.~Zhan.
\newblock Formal verification of quantum algorithms using {Q}uantum {H}oare
  {L}ogic.
\newblock In I.~Dillig and S.~Tasiran, editors, {\em Computer Aided
  Verification}, pages 187--207, USA, 2019. Springer International Publishing.

\bibitem{QHLProver}
J.~Liu, B.~Zhan, S.~Wang, S.~Ying, T.~Liu, Y.~Li, M.~Ying, and N.~Zhan.
\newblock Quantum hoare logic.
\newblock {\em Archive of Formal Proofs}, March 2019.
\newblock \url{https://isa-afp.org/entries/QHLProver.html}, Formal proof
  development.

\bibitem{Mermin93}
N.~D. Mermin.
\newblock Hidden variables and the two theorems of john bell.
\newblock {\em Rev. Mod. Phys.}, 65:803--815, Jul 1993.

\bibitem{nielsen-book}
M.~A. Nielsen and I.~L. Chuang.
\newblock {\em Quantum Computation and Quantum Information: 10th Anniversary
  Edition}.
\newblock Cambridge University Press, USA, 10th edition, 2011.

\bibitem{ConcreteSemantics}
T.~Nipkow and G.~Klein.
\newblock {\em Concrete Semantics: With Isabelle/HOL}.
\newblock Springer Publishing Company, Incorporated, USA, 2014.

\bibitem{Coq_Quantum}
R.~Rand, J.~Paykin, and S.~Zdancewic.
\newblock Qwire practice: Formal verification of quantum circuits in {C}oq.
\newblock {\em Electronic Proceedings in Theoretical Computer Science},
  266:119--132, 02 2018.

\bibitem{scarani2019bell}
V.~Scarani.
\newblock {\em Bell {N}onlocality}.
\newblock Oxford Graduate Texts. Oxford University Press, Oxford, Aug. 2019.

\bibitem{mathQuantum}
W.~Scherer.
\newblock {\em Mathematics of Quantum Computing: An Introduction}.
\newblock Springer International Publishing, Switzerland, 01 2019.

\bibitem{loophole2023}
S.~Storz, J.~Sch\"ar, A.~Kulikov, P.~Magnard, P.~Kurpiers, J.~L\"utolf,
  T.~Walter, A.~Copetudo, K.~Reuer, A.~Akin, et~al.
\newblock Loophole-free {B}ell inequality violation with superconducting
  circuits.
\newblock {\em Nature}, 617:265--270, 2023.

\bibitem{jordan_afp}
R.~Thiemann and A.~Yamada.
\newblock Matrices, {J}ordan normal forms, and spectral radius theory.
\newblock {\em Archive of Formal Proofs}, August 2015.
\newblock \url{https://isa-afp.org/entries/Jordan_Normal_Form.html}, Formal
  proof development.

\bibitem{jordan}
R.~Thiemann and A.~Yamada.
\newblock Formalizing {J}ordan normal forms in {I}sabelle/{HOL}.
\newblock In J.~Avigad and A.~Chlipala, editors, {\em Proceedings of the 5th
  {ACM} {SIGPLAN} Conference on Certified Programs and Proofs, Saint
  Petersburg, FL, USA, January 20-22, 2016}, pages 88--99, USA, 2016. {ACM}.

\bibitem{Unruh_2019}
D.~Unruh.
\newblock Quantum relational hoare logic.
\newblock {\em Proceedings of the ACM on Programming Languages}, 3(POPL):1--31,
  Jan 2019.

\bibitem{vidick2016three}
T.~Vidick.
\newblock Three-player entangled {XOR} games are {NP}-hard to approximate.
\newblock {\em SIAM Journal on Computing}, 45(3):1007--1063, 2016.

\end{thebibliography}
\bibliographystyle{abbrv}
\end{document}